# Searching for evidence of strengthening from short-range order in the CrCoNi medium entropy alloy

Novin Rasooli Matthew Daly*

*Department of Civil, Materials, and Environmental Engineering, University of Illinois Chicago – 842 W. Taylor St., 2095 ERF (MC 246), Chicago, IL, 60607, United States*

**ABSTRACT**

The coupling of strength to short-range order (SRO) in the CrCoNi medium entropy alloy remains actively investigated, with conflicting reports supporting and opposing SRO-induced strengthening continuing to emerge. A direct observation of this effect is elusive, due to difficulties in the quantification of SRO. Here, we deliver a structurally agnostic analysis that instead searches for unusual patterns in crystal size effects as evidence of SRO-induced strengthening. For this purpose, we assemble a large dataset of strengthening measurements drawn from a range of thermomechanical processing conditions known to produce SRO. Based on a comparative analysis with pure metal benchmarks, we find no evidence for significant coupling of SRO to strengthening in CrCoNi, and that patterns suggesting a positive finding may be explained by cross-study measurement scatter. Nevertheless, we leverage our analysis to provide an upper bound estimate of SRO-induced strengthening in the unlikely scenario where other sources of scatter are negligible.

**Keywords:** short-range ordering; medium entropy alloys; Hall-Petch effect; twinning; yield phenomena

*Corresponding author: mattdaly@uic.edu (M. Daly)

The role of short-range order (SRO) in the dislocation-mediated strengthening of concentrated alloys (e.g., medium- and high entropy alloys, M/HEAs) remains the focus of significant activity within the metallurgy community. Here, the equimolar CrCoNi MEA serves as perhaps the most notorious example, with greatly varying reports on the importance of SRO-induced strengthening populating the literature. For instance, a recent theory designed to predict the strength of randomly-arranged M/HEAs finds good agreement with experimental measurements of yielding in CrCoNi [1,2]. Here, the contributions of solute misfit to strength are found to reduce under SRO relative to random arrangement, which counteracts athermal strengthening introduced by the disruption of SRO during slip [3]. These theoretical findings agree with recent experimental efforts that explore a range of heat treatments designed to explicitly encourage SRO formation, showing negligible effects on yield strengths [4–6]. These measurements stand in contrast to the seminal study of SRO-induced strengthening [7], where a ~25% increase in the yield strength is measured in a slowly cooled sample (presumably with SRO) compared to a water quenched reference (presumably randomly arranged). This early measurement is supported by more recent molecular dynamics (MD) simulations, which observe significant changes to Peierls barriers under SRO [8,9]. However, these efforts rely on an interatomic potential that has been shown to over/underestimate pairwise interactions in CrCoNi [10]. It is therefore unclear if these studies measure an SRO-induced strengthening that is thermodynamically accessible. An authoritative conclusion on SRO-induced strengthening is further complicated by controversies in the measurement of local ordering. Most notably, the attribution of diffuse intensities in electron diffraction patterns to SRO formation (common in CrCoNi studies [5,7,11]) has been challenged, with symmetry breaking lattice features (e.g., static and thermal lattice displacements, and planar defects) providing an alternate explanation for extra reflections [12]. Although understanding of the thermodynamic conditions for SRO formation appears to be converging [13,14], its coupling to strength remains of great interest and efforts to explore SRO effects on strengthening persist. Indeed, the past months offer several high visibility studies both supporting [13,15] and opposing [6] SRO-induced strengthening in CrCoNi.

Here, we analyze the evidence for SRO-induced strengthening in the CrCoNi MEA through examination of trends in crystal size effects. In this analysis, we combine existing datasets with our own measurements that are targeted to capture strengthening in CrCoNi microstructures not well-represented in the literature. Collectively, this data represents a widely varying range of thermomechanical processing conditions, from which differing levels of SRO are anticipated to arise within and across datasets. In contrast to previous studies that focus on the assessment of induced strengthening by direct measurement of SRO, we present an ordering agnostic perspective that instead searches for unusual strengthening patterns that cannot be explained by other mechanisms. This approach is inspired in part by the size effect considerations presented in portions of Yin et al. [2], and offers the distinct advantage of assessing a large number of CrCoNi microstructures (albeit coarsely) without requiring direct measurement of SRO. To the authors' best knowledge, this study contemplates the largest dataset compiled to date for CrCoNi, with many measurements drawn from processing conditions now known to produce SRO.

We begin our analysis with a presentation of the single parameter Hall-Petch model, which is routinely applied to capture size effects in CrCoNi:

$$\sigma_y = \sigma_o + \frac{k}{\sqrt{D}}, D \in \{d, t\} \tag{1}$$

where $\sigma_y$ is the yield stress, $\sigma_o$ is the friction stress, $k$ is the strengthening coefficient, and $D$ is the crystal size. The CrCoNi system is known to twin profusely. Consequently, the literature diverges in its reporting of crystal size, with measurements excluding twin boundaries ($d$) and including twin boundaries ($t$) both prevalent. Figure 1a provides a summary of the Hall-Petch model parameters we have collected for the CrCoNi MEA. These parameters are also tabulated in the Supplementary Materials (Table S1). As shown in the figure, there are significant differences in the reported strengthening coefficients and friction stresses, which range between 265 – 598 MPa.μm$^{1/2}$ and 135 – 230 MPa (for $D = t$), and 537 – 815 MPa.μm$^{1/2}$ and 143 – 218 MPa (for $D = d$), respectively. To contextualize the friction stress values, the range of literature strengths for single crystals (i.e., $D \rightarrow \infty$) is marked by red-dashed lines (178 – 211 MPa) and the study-weighted average is drawn in red-dotted stroke (200 MPa). Here, single crystal strengths have been calculated from measurements of critical resolved shear stresses adjusted by a Taylor factor of 3.06 [4,16,17]. Literature values for single crystal measurements are tabulated in the Supplementary Materials (Table S2). Figure 1b provides two examples of measurement scatter in size effects when twins are excluded (black stroke) and included (blue stroke) from the crystal size, and Figure 1c compares two studies where crystal sizes are reported using the same size effect parameter (including twin boundaries). The dashed ellipses indicate crystal sizes where yield strengths deviate from Hall-Petch fitting, and significant differences between Hall-Petch models are evident in both examples. Here, unaccounted twinning effects offer one potential explanation for intra-study deviations [18]. Furthermore, cross-study differences (or agreement) in Hall-Petch parameters have been argued as a signature of SRO-induced strengthening (or lack thereof) [2,13], due to the varying thermomechanical treatments required to refine CrCoNi microstructures for size effect comparisons. We present these data here to contemplate differences in the definitions of crystal size and other microstructural sources of strengthening (e.g., twin boundaries) in the assessment of SRO effects. Our examination of the collected data leads to the following key observations: 1) The largest cross-study deviations in Hall-Petch parameters (Figure 1a) appear to result from differences in reporting of crystal sizes (i.e., whether excluding or including twin boundaries). However, SRO-induced strengthening cannot be immediately excluded, where the apparently large deviations between Hall-Petch models in studies using the same definition of crystal size (e.g., Figure 1c) motivate further analysis; 2) Strengthening due to twinning is governed by multiple microstructural parameters (e.g., twin spacing, twin thickness, and twin fraction [19–21]). Consequently, nonlinearities in Hall-Petch models may arise from differences in twin features that are not adequately captured by a single parameter model, even in studies where twins are included in the crystal size.

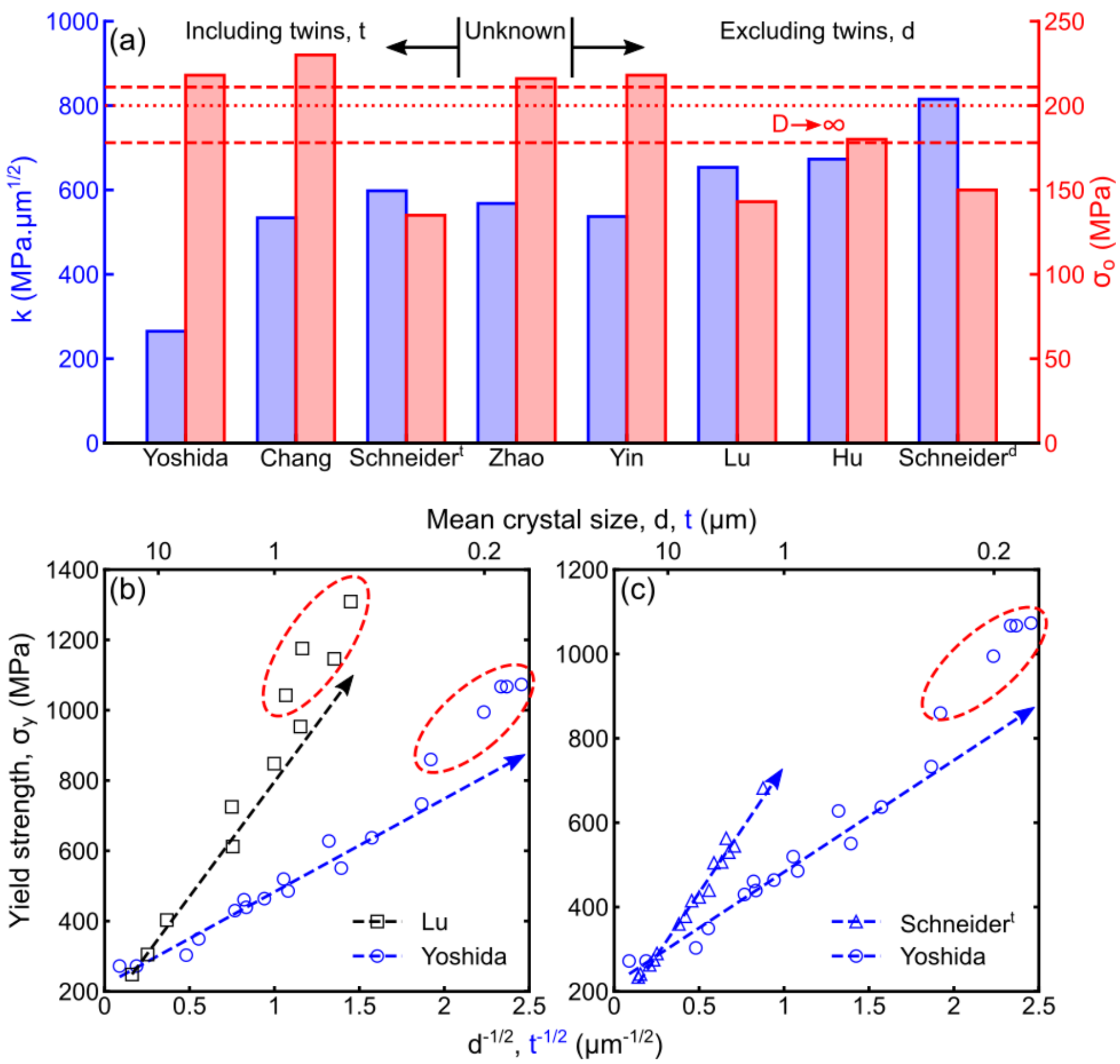

**Figure 1:** (a) Hall-Petch strengthening coefficients ($k$) and friction stresses ($\sigma_o$) for CrCoNi available in the literature [2,18,22–27]. Crystal sizes are reported including ($t$) and excluding ($d$) twin boundaries and are separated on the left and right sides of the panel, respectively. The crystal size representation reported in Zhao et al. [26] is unknown. Schneider et al. [24] provide both crystal size measurements, which are differentiated by the superscript. Literature ranges for single crystal strength measurements ($D \rightarrow \infty$) are provided in red-dashed stroke [4,16,17], with the study-weighted average (200 MPa) given by the red dotted line. (b) Comparison of size effects across two studies that differ in their measurement of crystal sizes [18,22]. Blue and black stroke mark crystal sizes reported including and excluding twin boundaries, respectively. (c) Variations in size effects between two studies that report crystal size including twin boundaries [22,24]. Deviations from Hall-Petch models are marked by red ellipses in (b) and (c).

To explore these observations, we have searched the available size effect literature for the CrCoNi MEA to analyze broader patterns in the scatter of measured strengths across varying processing conditions. In this search, we have selected studies where crystal size measurements include twin boundaries or where this representation could be measured from the published figures. As a part of this dataset, we report the structural parameters of twinned microstructures, where available [22,28–30]. The combined microstructural parameters from these sources are plotted in Figure 2a, and the corresponding numerical values are provided in the Supplementary Materials (Table S3). We have also synthesized and measured the yield strengths of an additional set of CrCoNi samples to access twinned microstructures not well-represented in the literature. Here, an ingot of equimolar CrCoNi (elemental feedstocks > 99.9% purity) was synthesized by vacuum arc remelting, following methods reported in our previous work [31,32]. The as-cast ingot was then hot-rolled at 900 °C to a ~30% thickness reduction and water quenched. The rolled sample was homogenized at 1200 °C for 12 hours in an alumina tube furnace under argon protection, and then water quenched. To obtain varying twinned microstructures, the homogenized sample underwent cold rolling to an additional 50% thickness reduction, was subdivided, and then aged at different temperatures (1123 – 1473 K) and holding times (5 – 120 min), followed by water quenching. X-ray diffraction (XRD) experiments were performed in a reflection geometry using a Bruker D8 Discover diffractometer, covering a 2θ range of 30–100° with 0.1° increments and a 1 s dwell time, under Cu Kα radiation (0.1541 nm X-ray wavelength). The results of XRD peak indexing confirm the anticipated single-phase face-centered cubic structure with a lattice parameter of 0.356 nm (Figure 2b), in good agreement with the literature [5,33]. Energy dispersive X-ray spectroscopy further confirmed an equimolar composition without evidence of microscale segregation (not shown). To measure the microstructure parameters of our samples, we performed scanning electron microscopy (SEM) with electron backscatter diffraction (EBSD) analysis. This analysis was conducted using a JEOL-IT500HR instrument operating at 20 kV with EBSD step sizes of 0.3–2 μm, following practices outlined by Anthony and Miller [34]. Crystals were segmented using a 5° misorientation criterion using the MTEX toolbox [35] (Figure 2c). Combined with the literature dataset, we report size effects over a range of microstructures measuring crystal sizes of 0.17 – 129 μm (including twin boundaries), twin thicknesses of 0.2 – 8.7 μm, twin spacings of 0.2 – 11.1 μm, and twin fractions of 0.30 – 0.48. Crystal sizes were measured using the linear intercept method as described in ASTM E112 [36]. Twin thicknesses are reported as a caliper measurement and twin fractions are calculated as the length of twin boundaries divided by the total crystal boundary length. Twin spacing measurements are based on the stereological relations described in Bouaziz and co-workers [19]. Details on this methodology are provided in the Supplementary Materials (Section 1). To assess the mechanical properties of our samples, we collected Vickers microhardness measurements under a 100 gf load and 20 s dwell time, with results averaged from at least five measurements per sample. To estimate the yield stress from Vickers microindentation measurements, we use a constraint factor of 5.55. This value is averaged from same-sample measurements of the Vickers hardness and uniaxial tensile yield strengths [5,28,37]. Although a constraint factor of 3 is often used to estimate yield strengths, this treatment is inaccurate for materials such as CrCoNi that exhibit significant work hardening [38]. We discuss our approach to obtain the constraint factor and its application to the Vickers hardness measurements collected in our lab in the Supplementary Materials (Section 2).

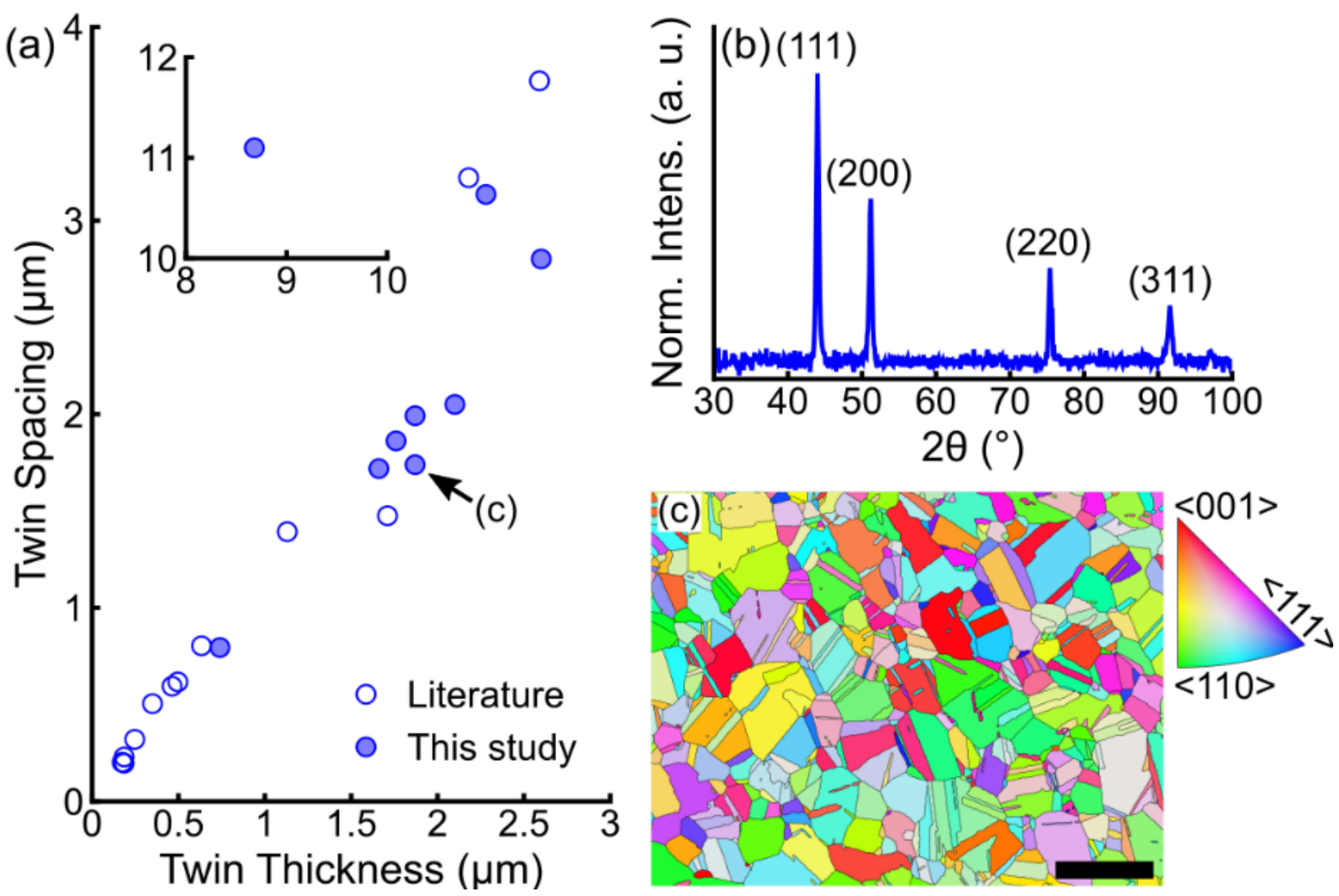

**Figure 2:** (a) The twin thickness and spacing for the CrCoNi microstructures examined in this study. Measurements from our samples are combined with literature data for reports where twin features were available [22,28–30]. The data point in the top-left is referenced against the inset axes. (b) The indexed XRD pattern for the homogenized ingot synthesized for this study. (c) A typical inverse pole figure map used to measure microstructure features in CrCoNi. The scale bar measures 30 μm.

Figure 3a provides the Hall-Petch plot for all collected CrCoNi measurements (74 data points across 15 independent studies). The complete dataset, including the sample processing conditions (where available), is detailed in the Supplementary Materials (Table S6). This data is plotted along with a best fit of Eq. (1). Here, we take $\sigma_o$ = 200 MPa from single crystal measurements, yielding a strengthening coefficient of $k$ = 387 MPa.μm$^{1/2}$ by least squares fitting, which falls in the mid-range of literature reports (Figure 1a). We further provide the 95% prediction interval for this dataset in dashed blue stroke, which estimates the scatter in additional measurements within two standard deviations. Given the relatively high coefficient of determination ($R^2$ = 0.84), a first interpretation supports the null finding. That is, SRO-induced strengthening, if present, has a very modest or perhaps negligible effect on yielding in CrCoNi [2–6]. To further explore this observation, we estimate the friction stress of individual measurements ($\hat{\sigma}_o$) by inversion of Eq. (1). We correlate these estimators with the aging temperatures, where available (Figure 3c), and compare against single crystal measurements. This information is stacked with a plot of pairwise correlations as defined by the Warren-Cowley (WC) parameter (Figure 3b), which matches aging temperatures with predictions of the equilibrium SRO structure (i.e., where WC $\neq$ 0). In this estimation, we inherently assume that SRO-induced strengthening contributes exclusively to lattice resistance (and not size effects) and attempt to correlate patterns in $\hat{\sigma}_o$ with aging conditions most favorable to SRO formation. Consequently, deviations from $\sigma_o$ = 200 MPa arise from a combination of SRO-induced strengthening (or weakening), unaccounted multiparameter twinning effects, and measurement scatter. Interestingly, the largest deviations and scatter arrive at above 973 K where the degree of SRO formation is reduced. Slower kinetics in the 573 – 873 K range may be largely discounted as some samples were aged for up to 500 h. In an effort to rationalize these observations, we examine the effects of cross-study measurement scatter by comparing the number of independent studies available for each aging condition (secondary axis, Figure 3c). As shown in the figure, the measurement scatter appears to be highly correlated with the number of independent reports, which favors a measurement error interpretation.

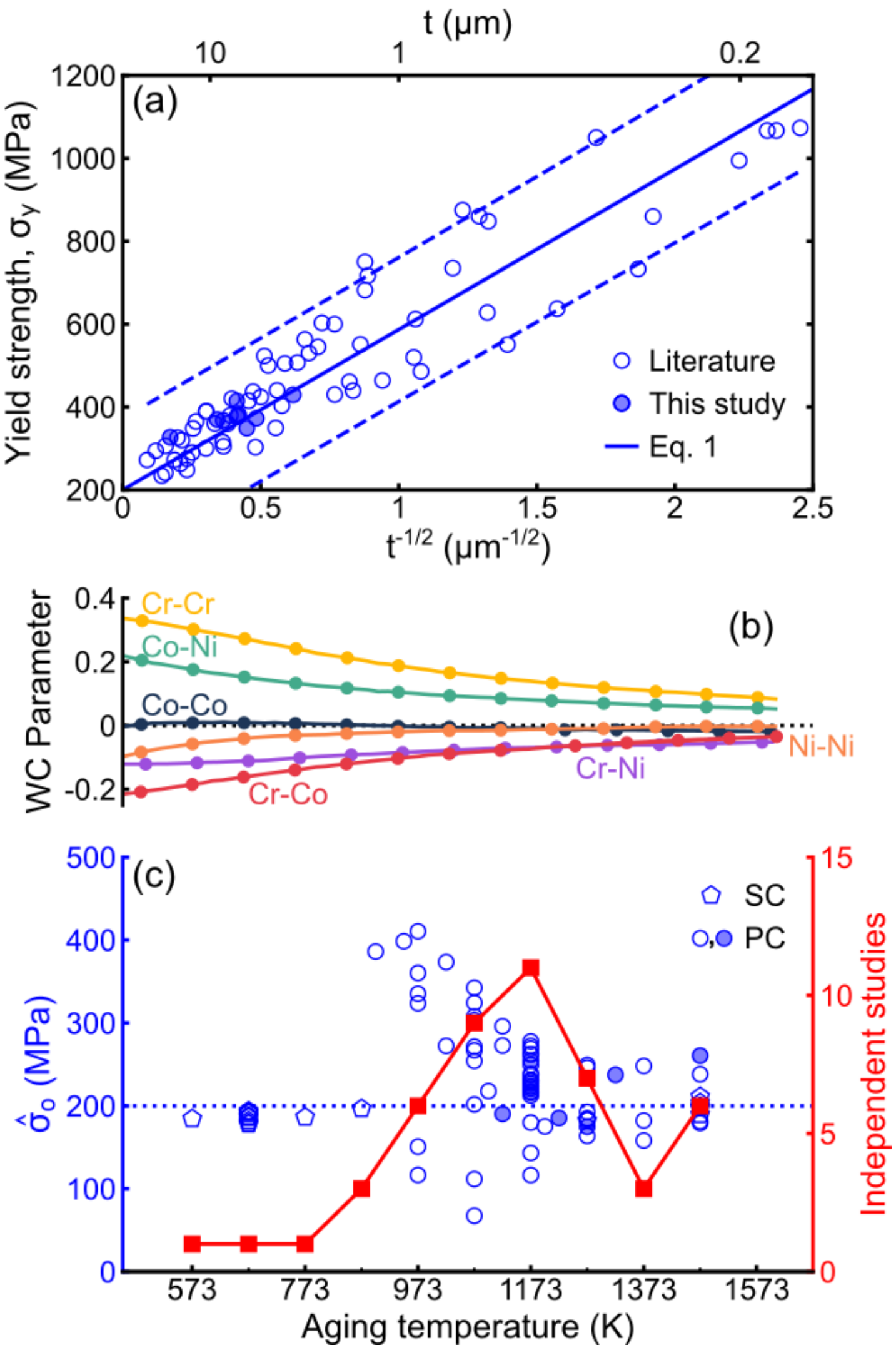

**Figure 3:** (a) Hall-Petch plot of 74 measurements from 15 independent studies of CrCoNi strengthening. Data produced in this study is separately marked. Literature data is obtained from Refs. [18,22,23,25,27–29,33,39–44]. The 95% prediction intervals are provided in dashed blue stroke. (b) The Warren-Cowley (WC) parameter for pairwise correlations in CrCoNi after Freitas and co-workers [10,45]. (c) Estimates of the friction stress ($\hat{\sigma}_o$) from inversion of Eq. (1) with aging temperature. Polycrystalline (PC) datapoints are sourced from our samples (filled markers) and the literature references (empty markers). Individual single crystal (SC) measurements are also provided [4,16,17], with the study weighted average of 200 MPa marked in blue dotted stroke. The aging data is binned in increments of 100 K and the number of independent studies for each bin is plotted in red stroke on the secondary axis.

Although a further quantitative deconvolution of individual contributors to scatter within the CrCoNi dataset is not possible, we draw insights on their likely effects by comparison with benchmark systems. For this purpose, we compare Hall-Petch models of CrCoNi with pure Cu and Ni (Figure 4a). The former is selected as a benchmark for twinned microstructures and the latter as a face-centered cubic parent element of the MEA. Here, we emphasize that SRO (and solute strengthening) do not contribute to measurement scatter in Cu and Ni. Rather, we offer these systems as baselines for unaccounted twinning and measurement scatter in similar size effect studies. Measurements for the pure metals were obtained from an aggregated dataset that sources multiple independent studies [46], with values drawn from crystal sizes in the range of ~100 nm – 500 μm. Examination of the prediction intervals (Figure 4a) shows that CrCoNi exhibits the narrowest scatter and lowest root mean squared error (RMSE) of 87 MPa, compared to 101 and 232 MPa for Cu and Ni, respectively (Figure 4b). From this comparison, we note the following observations: 1) The larger scatters in pure metal datasets undermine significant SRO-induced strengthening in CrCoNi; 2) Twin boundary strengthening is adequately captured by a single parameter model with accuracies comparable to a Cu benchmark; and 3) The large scatter in the Ni dataset, combined with correlations in the number of independent CrCoNi studies (Figure 3c), suggest cross-study measurement error as the dominant contributor to scatter. Consequently, we conclude that there is currently no evidence for a significant coupling of SRO to strengthening in the CrCoNi MEA, despite a well-explored microstructure and thermomechanical processing space. That is, although SRO likely exists in the measured samples, it does not appear to have a significant effect on yielding, which supports earlier smaller-data reports of negligible strengthening [2–6]. Additionally, the scatter observed in literature can be largely attributed to cross-study measurement error after controlling for differences in definition of the crystal size parameter. Nevertheless, if other sources of scatter are neglected, we offer 87 MPa (i.e., the RMSE) as an upper bound for the expected value of strengthening achieved by SRO in CrCoNi. We acknowledge that this finding assumes a Gaussian distribution for SRO-induced strengthening/weakening that is independent of crystal size, which is difficult to verify in this aggregated dataset. While we stress that the true expected value of SRO-induced strengthening (if any) is likely much lower than 87 MPa, we believe this upper bound is useful in calibrating community efforts, particularly given the challenges associated with quantifying SRO and directly correlating its presence with mechanical behavior. Furthermore, although we do not find evidence for significant SRO-induced strengthening in CrCoNi, this interpretation does not preclude coupling in other MEA/HEAs, where high quality strengthening studies remain underdeveloped.

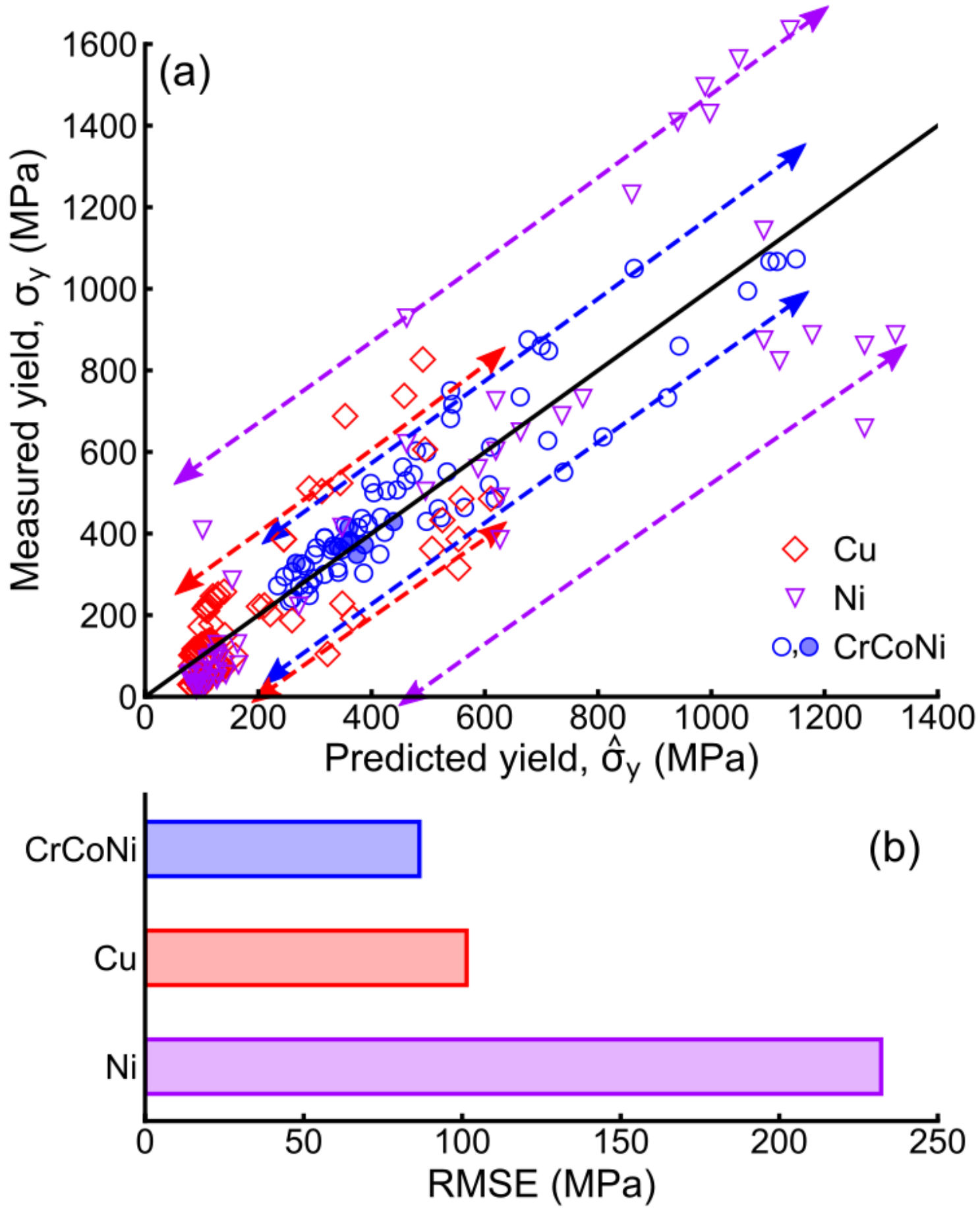

**Figure 4:** (a) The measured yield strengths ($\sigma_y$) compared to predicted yields ($\hat{\sigma}_y$) by Hall-Petch models of CrCoNi (this work), Cu, and Ni. Our measurements for CrCoNi are noted with filled markers and literature values appear as empty markers [18,22,23,25,27–29,33,39–44]. Data for Cu and Ni are obtained from Cordero et al. [46]. The 95% prediction intervals are provided in dashed stroke. The line of zero measurement error is drawn in solid black stroke. (b) The root mean squared error (RMSE) for the Hall-Petch models of each system.

**ACKNOWLEDGEMENTS**

This material is based upon work supported by the National Science Foundation under Grant No. 2144451. The authors would like to express their gratitude to Dr. Shuhei Yoshida, Dr. Praveen Sathiyamoorthi, Dr. Jiashi Miao, and Dr. Supriyo Chakraborty discussions and in many cases for providing raw EBSD data on aged CrCoNi samples, which helped populate the twinned microstructure parameter space. These datasets appear in peer-reviewed works in Refs. [22,28–30]. We also thank Professor Victoria Miller and her research group for helpful discussions on the analysis of EBSD datasets of twinned microstructures.

**AUTHOR CONTRIBUTIONS**

Novin Rasooli: Formal Analysis; Data Curation; Visualization; Writing – Original Draft; Writing – Review & Editing. Matthew Daly: Conceptualization; Formal Analysis; Funding Acquisition; Project Administration; Supervision; Visualization; Writing – Original Draft; Writing – Review & Editing.

**COMPETING INTERESTS**

There are no conflicts to declare.